\documentclass[aps,pra,superscriptaddress,floatfix,showpacs,10pt, twocolumn,showkeys]{revtex4}
\usepackage{bbm}
\usepackage{amsfonts}
\usepackage[dvips]{graphicx}
\usepackage{amsmath,amsfonts,amssymb,graphics,graphicx,epsfig,color,times,bbm}

\begin{document}

\title{Nonlocality Distillation for High-Dimensional System}

\author{Guo-Zhu Pan}
\affiliation{Key Laboratory of Opto-electronic Information
Acquisition and Manipulation, Ministry of Education, School of
Physics {\&} Material Science, Anhui University, Hefei 230039,
People's Republic of China.} \affiliation{School of Material and
Chemical Engineering, West Anhui University, Lu'an 237012, People's
Republic of China.}

\author{Chao Li}
\affiliation{Key Laboratory of Opto-electronic Information
Acquisition and Manipulation, Ministry of Education, School of
Physics {\&} Material Science, Anhui University, Hefei 230039,
People's Republic of China.}

\author{Zheng-Gen Chen}
\affiliation{Key Laboratory of Opto-electronic Information
Acquisition and Manipulation, Ministry of Education, School of
Physics {\&} Material Science, Anhui University, Hefei 230039,
People's Republic of China.}

\author{Ming Yang \footnote{Corresponding Author: mingyang@ahu.edu.cn}}
\affiliation{Key Laboratory of Opto-electronic
Information Acquisition and Manipulation, Ministry of Education,
School of Physics {\&} Material Science, Anhui University, Hefei
230039, People's Republic of China.}

\author{Zhuo-Liang Cao}
\affiliation{School of Electronic information Engineering, Hefei
Normal University, Hefei 230061, People's Republic of China.}

\begin{abstract}
The intriguing and powerful capability of nonlocality in
communication field ignites the research of the nonlocality
distillation. The first protocol presented in Ref[Phys. Rev. Lett.
102, 120401] shows that the nonlocality of bipartite binary-input
and binary-output nonsignaling correlated boxes could be amplified
by 'wiring' two copies of weaker-nonlocality boxes. Several
optimized distillation protocols were presented later for bipartite
binary-input and binary-output nonsignaling correlated boxes. In
this paper, we focus on the bipartite binary-input and
multi-nary-output nonsignaling correlated boxes---high-dimensional
boxes, and design comparators-based protocols to achieve the
distillation of high-dimensional nonlocality. The results show that
the high-dimensional nonlocality can be distilled in different ways,
and we find that the efficiencies of the protocols are influenced
not only by the wirings but also by the classes the initial
nonlocality boxes belongs to. Here, the initial nonlcalities may
have the same violation of the high-dimensional Bell-type
inequality, but they can fall into different classes, which shows
that the value of the violation of the high-dimensional Bell-type
inequality is not the only representation of nonlocality, and the
combination manner(classes) in the expression of the correlated
boxes is another important nonlocal representation too. The current
protocols are compatible with the previous two-dimensional
nonlocality distillation protocols, but our protocols are more
powerful and universal than previous ones in the sense that the
current protocols can be applied to the system with any dimension
rather than the only two-dimension system in the previous protocols.
\end{abstract}

\pacs{03.65.Ud,03.67.Mn}
\date{\today}
\keywords{Bell-type inequalities, Nonlocality distillation,
High-dimensional nonsignaling correlated boxes} \maketitle

\section{introduction}
The result of the measurements on spatially separated maximal
entangled state runs counter to local realism, which was named as
quantum nonlocality. Quantum nonlocality was the focus of the query
of quantum mechanics (QM) in the original time when QM was
born\cite{EPR}. After Bell\cite{Bell} gave the first inequality in
his theory which classical correlation must obey but QM doesn't,
quantum nonlocality could be directly perceived through the
violation of the corresponding Bell-type inequality\cite{Bell-type}.

Tsirelson\cite{Tsirelson}'s study shows that quantum correlations
are limited by $B_Q=2\sqrt2$ under the
Clauser-Horne-Shimony-Holt\cite{CHSH} (CHSH) expression of Bell's
theory. Popescu and Rohrlich\cite{PR} (PR) presented an unnatural
correlation which violates CHSH inequality by its algebraic maximum
4. All these results illustrate that the set of quantum correlations
is bigger than the set of local correlations bounded by Bell-type
inequalities, but it is still included in a bigger set of the
general nonsignaling correlations, which can reach the algebraic
maximal violation of Bell-type inequalities\cite{resource}.

As an information-theoretic resource, the general nonsignaling
models in the form of nonsignaling boxes were presented for
convenient study of the set of different
correlations\cite{resource}. The nonsignaling boxes not only can
give simulations of quantum nonlocal correlations (QNC) but also can
indicate the correlations beyond QNC---postquantum correlations
(PQC), as PR correlations for example. Some extraordinary
information processing abilities are revealed by the study of
postquantum theories. The PQC can make communication complexity
trivial\cite{vDam,trivial,BS}, simulate quantum entanglement without
communication\cite{simulate,simulate2}, and make dynamics rich in
the process of nonlocality swapping\cite{NLS}. 'Nonlocal
computation' is allowed by the PQC\cite{Linden}, and 'information
causality' is also violated\cite{Pawlowski}.

Generally, The stronger nonlocality the resources have, the more
useful they are. The question on how to obtain more nonlocality from
weak ones is naturally raised to us. Foster et.al\cite{FWW} gave an
affirmative answer to this question by presenting a deterministic
nonlocality distillation protocol (FWW) for bipartite binary-input
and binary-output correlated nonlocal boxes. Then Brunner
et.al\cite{BS} made an improved protocol (BS) where one box's
outputs are added into another box's inputs. If these two boxes are
identical, this kind of wiring can be regarded as an kind feedback
in this sense. Later Allcock \emph{et al}\cite{ABLPSV} gave a more
efficient distillation protocol (ABLPSV) by changing the nonlocal
elements, and H{\o}yer et.al\cite{HR} made a 'depth-3' method (HR)
by using three boxes in each distillation round. All these protocols
are working fine for the bipartite correlated nonlocal boxes in
certain scope. Recently, the distillation protocol for multipartite
nonlocality was discussed, where Li-Yi Hsu and Keng-ShuoWu\cite{HW}
generalized the bipartite FWW protocol into the $n$-partite case.
Brunner et.al\cite{Activation} discussed the bound nonlocality and
activation of distillabe nonlocality under the
Elitzur-Popescu-Rohrlich\cite{EPR2} decomposition.
Forster\cite{BoundD} gave a way to bound the distillable nonlocality
of a resource by solving a related optimization problem.

Binary-outputs of boxes means that the boxes is two dimensional
ones, and so far, all the above-mentioned distillation protocols
only work for the two dimensional boxes. Here we will present a
novel nonlocality distillation protocol for higher dimensional
boxes. In our protocol, the boxes to be distilled is replaced by
arbitrary high-dimensional boxes and new wiring method is applied on
it. The results show that our protocol distills different boxes with
different efficiencies, and the protocol also works for the more
general 'noisy' correlated nonlocal boxes.

\section{bipartite high-dimensional nonsignaling correlated boxes}\label{s2}
\begin{figure}[htbp]
\includegraphics[width=0.40\textwidth]{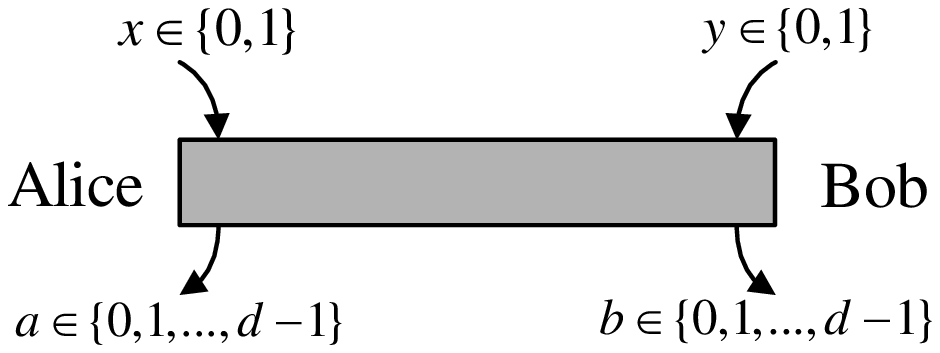}
\caption{\label{hdbox}The $d$-dimensional bipartite nonsignaling
correlated box. The two inputs of the box are binary and the two
outputs are $d$-nary.}
\end{figure}

QM is nonlocal and the nonsignaling models can give simulations of
the nonlocal correlations from the respective measurements on
spatially separated quantum states. As shown in Fig.\ref{hdbox}, a
nonsignaling model is a black box which has two input terminals and
two output terminals on the two sides belonging to two spatially
separated users-Alice and Bob. The inputs $x$ and $y$ simulate Alice
and Bob's possible measurements in quantum measuring process, and
the outputs $a$ and $b$ simulate the possible outcomes after Alice
and Bob's measurements. Each user can carry out two possible
measurements when the inputs $x$ and $y$ are both binary. The fact
that the outputs $a$ and $b$ can be $0,1,...,d-1$ means that each
measurement of their two choices may have $d$ different outcomes,
i.e., the system is $d$-dimensional. In the generalized nonsignaling
theories, when $d=2$, the box is a two-dimensional one, i.e., the
one being widely discussed before. $d>2$ means the box is a
$d$-dimensional one, which is in accordance with $d$-dimensional
quantum system.

The joint probability distribution $P(ab|xy)$ characterizes the
nonlocality of the box and the corresponding quantum system the box
simulates. The nonlocality of the box is directly reflected by the
violation of the high-dimensional Bell inequality---the
$d$-dimensional generalization of CHSH inequality (CGLMP), which was
developed in Ref.\cite{CGLMP}. We adopt the scalar product-type
expression in Ref.\cite{NLS} and denote the box's CGLMP value as
$\Vec{\text{CGLMP}}\cdot\Vec{P}(ab|xy)$. For a $d$-dimensional
system, the inequality has the form:
\begin{equation}
\Vec{\text{CGLMP}}\cdot\Vec{P}(ab|xy)=E_{xy}+E_{\bar{x}y}+E_{x\bar{y}}-E_{\bar{x}\bar{y}}\leq2,
\end{equation}
where $\bar{x}$ and $\bar{y}$ indicate bit flips, and $E_{xy}$ is
defined as
\begin{equation}
\begin{aligned}
E_{xy}=&\sum\limits_{k=0}^{d/2-1}(1-\frac{2k}{d-1})[P((b-a)\text{mod
}d=-k|xy)\\
&-P((b-a)\text{mod }d=k+1|xy)].
\end{aligned}
\end{equation}
The local bound is always 2 and the algebraic maximal violation of
CGLMP inequality is also 4 for arbitrary high-dimensional system.

If the inputs $x,y\in\{0,1\}$ and outputs $a,b\in\{0,1,...,d-1\}$ as
depicted in Fig.\ref{hdbox}, the nonlocal vertex is given by
\begin{equation}
P_{\text{NL}}^d(ab|xy)=
\begin{cases}
1/d,&\text{if }(b-a)\text{mod }d=xy,\\
0,&\text{otherwise}.
\end{cases}
\end{equation}
It has an extremal violation 4 of CGLMP inequality. When $d=2$, it
is the PR correlation.

The $d$-dimensional local correlated box is defined as:
\begin{equation}
P_{\text{Lc}}^d(ab|xy)=
\begin{cases}
1/d,&\text{if }(b-a)\text{mod }d=0,\\
0,&\text{otherwise}.
\end{cases}
\end{equation}
Its outputs are independent of the inputs and it violates the CGLMP
inequality by $2$. When $d=2$, it is the box '$P^{c}$' in the
earliest two distillation protocols\cite{FWW,BS} of $2$-dimensional
boxes.

The $d$-dimensional local deterministic box is described as:
\begin{equation}
P_{\text{Ld}}^d(ab|xy)=
\begin{cases}
1,&\text{if }a=d-1,b=d-1,\\
0,&\text{otherwise}.
\end{cases}
\end{equation}
It also violates CGLMP inequality by $2$ as
$P_{\text{Lc}}^d(ab|xy)$, and it is the box '$P_{\text{L}}^{0101}$'
appeared in the third protocol\cite{ABLPSV} when $d=2$.

The $d$-dimensional fully mixed box $\openone^d(ab|xy)=1/d^2,$
$\forall a,b,x,y$, and, for simplicity, it will be written as
$\openone$ hereafter.

\section{nonlocality distillation of $d$-dimensional boxes}\label{s3}

\subsection{A brief review of nonlocality distillation in bipartite two-dimensional
boxes}\label{s3A}

One nonsignaling correlated box with more nonlocality could be
generated by certain local operations (local classical operations
such as connecting users' respective inputs and outputs etc.) on
many ones without classical communication, which is the so-called
nonlcoality distillation. Refs.\cite{resource,BS,ABLPSV} pointed out
that the distillation protocol can be viewed as classical circuitry,
i.e., the operations on the inputs and outputs are considered as
\emph{wirings}, and the nonsignaling correlated boxes correspond to
\emph{components}.

In the earliest protocol, FWW protocol, two users Alice and Bob made
the wiring between two copies of box
$P_{\epsilon,\text{c}}^2={\epsilon}P_{\text{NL}}^2+(1-\epsilon)P_{\text{Lc}}^2$
by applying XOR wirings on their respective outputs. After the
protocol, the proportion of the nonlocal element becomes
$2\epsilon-2\epsilon^2$, and the protocol works well on the region
$0<\epsilon<1/2$ of $\epsilon\in[0,1]$. Later, BS protocol as an
excellent one which adds extra AND wirings on the second box's
inputs can always distill the same class of boxes when
$0<\epsilon<1$ with nonlocal proportion $\epsilon(3-\epsilon)/2$. In
ABLPSV protocol,  the components $P_{\epsilon,\text{c}}^2$ are
replaced by
$P_{\epsilon,\text{d}}^2={\epsilon}P_{\text{NL}}^2+(1-\epsilon)P_{\text{Ld}}^2$,
and the wirings changes correspondingly. After the distillation the
proportion of nonlocal element reaches the optimal
$2\epsilon-\epsilon^2$.

From the summary of previous protocols, we can find that the
efficiency of nonlocality distillation is decided not only by the
wirings but also by the type of the initial boxes. Here we replace
the binary-outputs boxes by $d$-nary-outputs ones, i.e., the
$d$-dimensional nonsignaling correlated boxes. The new
wirings---\emph{comparators} are presented to achieve the
distillation of high-dimensional boxes' nonlocality. Applying the
distillation protocol on different high-dimensional boxes, we find
that the distillation protocol has a selective efficiency to the
initial components.

\begin{figure}[tbp]
\includegraphics[scale=0.60,angle=0]{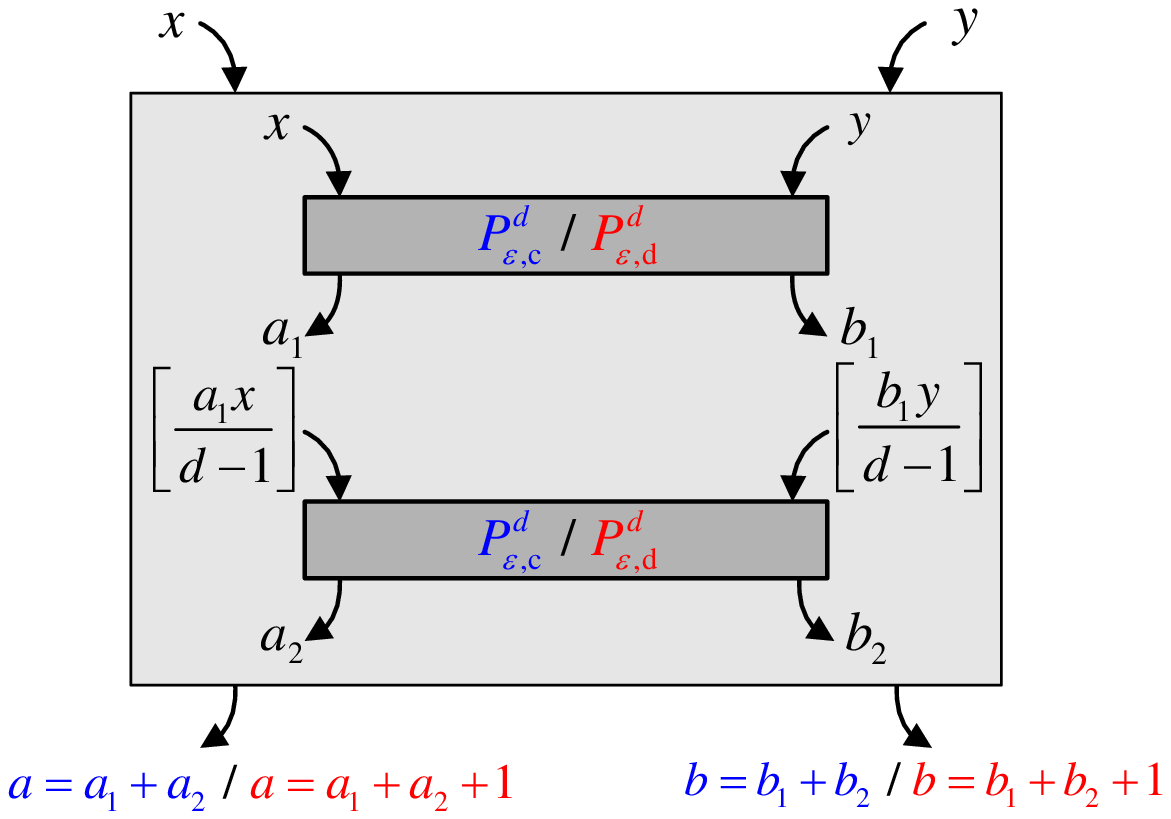}
\caption{\label{protocol}Distillation protocols of
$P_{\epsilon,\text{c}}^d$ and $P_{\epsilon,\text{d}}^d$. The color
of the wirings corresponding the same color of the boxes in the
figure. The second box's inputs are decided by the global inputs,
the first box's outputs and $d-1$, which is similar to a
\emph{comparator}. All the + in the wirings denote addition modulo
$d$. A slight distinction between the two protocols for different
boxes is on the global outputs.}
\end{figure}

\subsection{Protocol A: distillation of $P_{\epsilon,\text{c}}^d$}\label{s3B}
Consider the class of nonsignaling correlated boxes defined as
follows:
\begin{equation}\label{eq6}
P_{\epsilon,\text{c}}^d={\epsilon}P_{\text{NL}}^d+(1-\epsilon)P_{\text{Lc}}^d,
\end{equation}
where $0\leq\epsilon\leq1$ and $d>2$. The box
$P_{\epsilon,\text{c}}^d$ has a CGLMP value
$\Vec{\text{CGLMP}}\cdot\Vec{P}_{\epsilon,\text{c}}=2+2\epsilon$. As
shown in Fig.\ref{protocol} (ignore the illustration by red color),
Alice and Bob have two copies of $P_{\epsilon,\text{c}}^d$, and the
state before distillation is
\begin{equation}
\begin{aligned}
P_{\epsilon,\text{c}}^d P_{\epsilon,\text{c}}^d
=&\epsilon^2P_{\text{NL}}^d
P_{\text{NL}}^d+\epsilon(1-\epsilon)(P_{\text{NL}}^d
P_{\text{Lc}}^d+P_{\text{Lc}}^d
P_{\text{NL}}^d)\\&+(1-\epsilon)^2P_{\text{Lc}}^d P_{\text{Lc}}^d.
\end{aligned}
\end{equation}

The users proceed in each side as follows:
$x_1=x,x_2=\left[\dfrac{a_1}{d-1}\right]x,a=a_1+a_2,y_1=y,y_2=\left[\dfrac{b_1}{d-1}\right]y,b=b_1+b_2$,
where + in the wirings denotes the addition modulo $d$, and
$x_i,y_i,a_i,b_i$ denote the inputs and outputs bits of the $i$th
box. As the key of the protocol, $\left[\dfrac{k}{d-1}\right]$ is
given by
\begin{equation}
\left[\dfrac{k}{d-1}\right]=
\begin{cases}
1,&\text{if }k=d-1,\\
0,&\text{if }k<d-1.
\end{cases}
\end{equation}
which is similar to a \emph{comparator} in classical circuitry.

The wirings defined above take two initial boxes to one final box,
so there will be four cases of $P_i P'_i\rightarrow P_f$:

a).$P_{\text{NL}}^d P_{\text{NL}}^d\rightarrow P_{\text{NL}}^d$. We
can get $(b_1-a_1)\text{mod }d=xy$ from the first box, i.e.
$b_1=(xy+a_1)\text{mod }d$. For the second box we have
$(b_2-a_2)\text{mod
}d=\left[\dfrac{a_1}{d-1}\right]\left[\dfrac{b_1}{d-1}\right]xy=\left[\dfrac{a_1}{d-1}\right]\left[\dfrac{(xy+a_1)\text{mod
}d}{d-1}\right]xy=0$, so the global relation $(b-a)\text{mod
}d=[(b_1+b_2)\text{mod }d-(a_1+a_2)\text{mod }d]\text{mod
}d=[(b_1-a_1)\text{mod }d+(b_2-a_2)\text{mod }d]\text{mod }d=xy$.

b).$P_{\text{NL}}^d P_{\text{Lc}}^d\rightarrow P_{\text{NL}}^d$. For
the first box we have $(b_1-a_1)\text{mod }d=xy$, and for the second
box we have $(b_2-a_2)\text{mod }d=0$, so the final relation
$(b-a)\text{mod }d=xy$, too.

c).$P_{\text{Lc}}^d P_{\text{NL}}^d\rightarrow
\frac{1}{d}P_{\text{NL}}^d+(1-\frac{1}{d})P_{\text{Lc}}^d$. The
first box given by $(b_1-a_1)\text{mod }d=0$, implies $b_1=a_1$. For
the second box, $(b_2-a_2)\text{mod
}d=\left[\dfrac{a_1}{d-1}\right]\left[\dfrac{b_1}{d-1}\right]xy=\left[\dfrac{a_1}{d-1}\right]\left[\dfrac{a_1}{d-1}\right]xy$.
Finally, we get $(b-a)\text{mod
}d=\left[\dfrac{a_1}{d-1}\right]\left[\dfrac{a_1}{d-1}\right]xy$,
where $a_1$ is random. The final relation will be $(b-a)\text{mod
}d=xy$ when $a_1=d-1$ with probability $1/d$, and $(b-a)\text{mod
}d=0$ otherwise.

d).$P_{\text{Lc}}^d P_{\text{Lc}}^d\rightarrow P_{\text{Lc}}^d$.
Here $(b_1-a_1)\text{mod }d=0$ and $(b_2-a_2)\text{mod }d=0$, so we
have $(b-a)\text{mod }d=0$.

After the above four logical calculuses, the final state of the box
is given by
\begin{equation}
P_{\epsilon',\text{c}}^d=[(1+\frac{1}{d})\epsilon-\frac{1}{d}\epsilon^2]P_{\text{NL}}^d
+[1-(1+\frac{1}{d})\epsilon+\frac{1}{d}\epsilon^2]P_{\text{Lc}}^d,
\end{equation}
where $\epsilon'=(1+\frac{1}{d})\epsilon-\frac{1}{d}\epsilon^2$.
$\Vec{\text{CGLMP}}\cdot\Vec{P}_{\epsilon',\text{c}}^d-\Vec{\text{CGLMP}}\cdot\Vec{P}_{\epsilon,\text{c}}^d=2+2\epsilon'-(2+2\epsilon)>0$
is always tenable with $0<\epsilon<1$ and a finite dimension, which
implies the fraction of the nonlocal component is increased and the
distillation protocol succeeds.

\subsection{Protocol B: distillation of $P_{\epsilon,\text{d}}^d$}\label{s3C}
Suppose we use another superposition of different components as the
initial boxes, i.e.
\begin{equation}
P_{\epsilon,\text{d}}^d={\epsilon}P_{\text{NL}}^d+(1-\epsilon)P_{\text{Ld}}^d,
\end{equation}
where $0\leq\epsilon\leq1$ and $d>2$. Notice that this kind of
superposition has the same CGLMP value as that of the superpositions
in Eq.(\ref{eq6}) in protocol A. The initial state is
\begin{equation}
\begin{aligned}
P_{\epsilon,\text{d}}^d P_{\epsilon,\text{d}}^d
=&\epsilon^2P_{\text{NL}}^d
P_{\text{NL}}^d+\epsilon(1-\epsilon)(P_{\text{NL}}^d
P_{\text{Ld}}^d+P_{\text{Ld}}^d
P_{\text{NL}}^d)\\&+(1-\epsilon)^2P_{\text{Ld}}^d P_{\text{Ld}}^d.
\end{aligned}
\end{equation}

Through the wirings shown in Fig.\ref{protocol} (ignore the blue
part):
$x_1=x,x_2=\left[\dfrac{a_1}{d-1}\right]x,a=a_1+a_2+1,y_1=y,y_2=\left[\dfrac{b_1}{d-1}\right]y,b=b_1+b_2+1$,
we could get the following transformations: $P_{\text{NL}}^d
P_{\text{NL}}^d\rightarrow P_{\text{NL}}^d$, $P_{\text{NL}}^d
P_{\text{Ld}}^d\rightarrow P_{\text{NL}}^d$, $P_{\text{Ld}}^d
P_{\text{NL}}^d\rightarrow P_{\text{NL}}^d$, $P_{\text{Ld}}^d
P_{\text{Ld}}^d\rightarrow P_{\text{Ld}}^d$. The final box is
\begin{equation}
P_{\epsilon',\text{d}}^d=(2\epsilon-\epsilon^2)P_{\text{NL}}^d
+(1-2\epsilon+\epsilon^2)P_{\text{Ld}}^d.
\end{equation}
Obviously for $0<\epsilon<1$,
$\epsilon'=2\epsilon-\epsilon^2>\epsilon$, and the nonlocality of
the final box is bigger than the initial ones.

\begin{figure}[htbp]
\includegraphics[scale=0.75,angle=0]{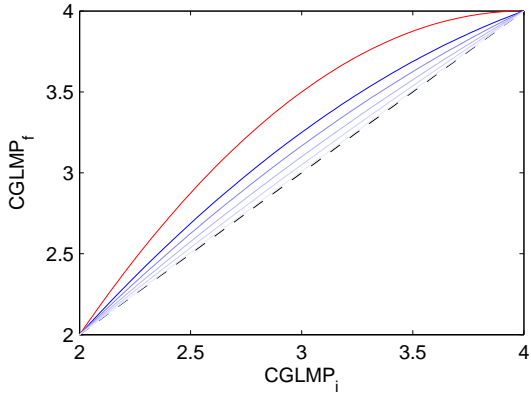}
\caption{\label{eff}Comparison of high-dimensional nonlocality
distillation protocols A and B. The final CGLMP values of the two
protocols are shown as the functions of the initial CGLMP value with
different colors in the figure. On the dashed line,
$\Vec{\text{CGLMP}}\cdot\Vec{P}_\text{f}=\Vec{\text{CGLMP}}\cdot\Vec{P}_\text{i}$.
The blue lines whose colors become brighter gradually describe the
distillation protocols A of $d=2,3,5,10,50$ successively. The red
line describes protocol B for any dimension $d$. In this figure, the
fact that the efficiency of protocol B is better than protocol A is
obvious.}
\end{figure}

The initial boxes in the two protocols have the same violation of
CGLMP inequality, but they induce different efficiency of
distillation in the similar wirings. The fact that protocol B has a
better efficiency than protocol A is revealed immediately from the
comparison shown in Fig.\ref{eff}.

It is easy to see that, our protocol A is compatible with the BS
protocol, and protocol B has an equal efficiency as the ABLPSV
protocol when $d=2$, so our protocol is effective for arbitrary
dimensional boxes. In the protocol A, the dimension of the boxes
appears in the coefficient of the final nonlocal fraction, that is
to say, the distillation efficiency is a function of the dimension
of the boxes (the protocol becomes trivial when
$d\rightarrow\infty$). But in the protocol B, the final box's
nonlocal fraction is not varying with the dimension of the boxes,
and is always bigger than protocol A.

\section{Nonlocality distillation in more general cases}\label{s4}

Our protocol can also work for the more general $d$-dimensional
nonlocal boxes, i.e. the noisy correlated boxes as such a ploytope:
\begin{equation}
P_{\xi,\gamma}^d=\xi P_{\text{NL}}^d+\gamma
P_{\text{Ld}}^d+(1-\xi-\gamma)\openone,
\end{equation}
where $\xi,\gamma\geq0$ and $\xi+\gamma\leq1$. $\openone$ is the
$d$-dimensional fully mixed box. After applying the wirings
presented in protocol B on two copies of the noisy box, and noticing
the extra transformations
$P_{\text{NL}}^d\openone\rightarrow\openone,
{\openone}P_{\text{NL}}^d\rightarrow\frac{1}{d^2}P_{\text{NL}}^d+\openone-\frac{1}{d^2}P_{\text{Lc}}^d,
P_{\text{Ld}}^d\openone\rightarrow\openone,
{\openone}P_{\text{Ld}}^d\rightarrow\openone,
\openone\openone\rightarrow\openone$, we can get the final box after
distillation:
\begin{equation}
\begin{aligned}
P_{\text{f}}^d&=[(1-\frac{1}{d^2})\xi^2+(2-\frac{1}{d^2})\xi\gamma+\frac{1}{d^2}\xi]P_{\text{NL}}^d+\gamma^2P_{\text{Ld}}^d\\
&+(1+\xi+\gamma)(1-\xi-\gamma)\openone-\frac{1}{d^2}\xi(1-\xi-\gamma)P_{\text{Lc}}^d,
\end{aligned}
\end{equation}

\begin{figure}[tbp]
\includegraphics[scale=0.70,angle=0]{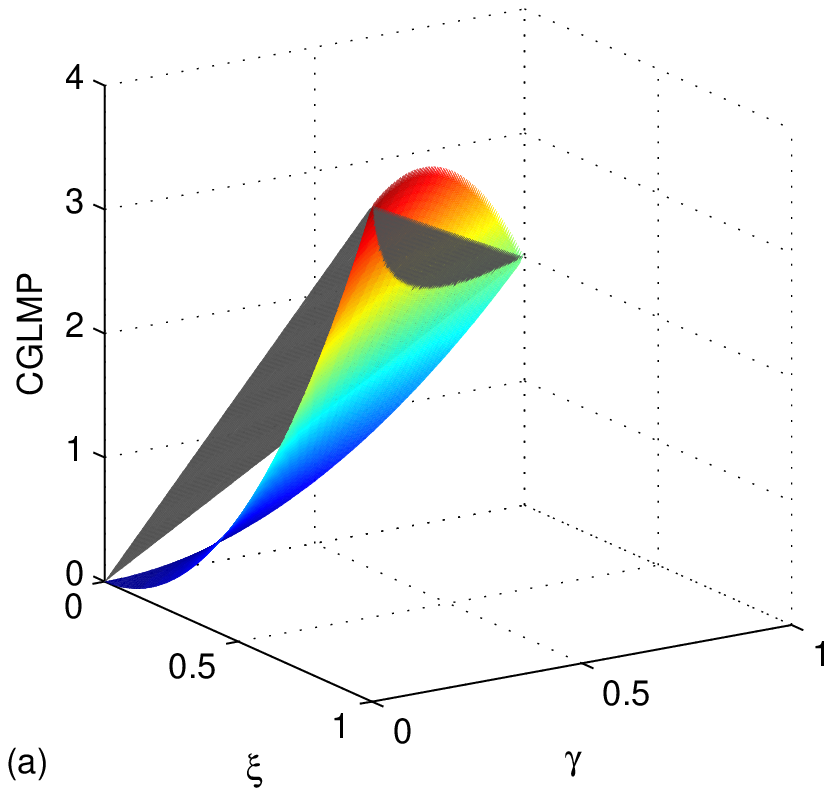}
\includegraphics[scale=0.70,angle=0]{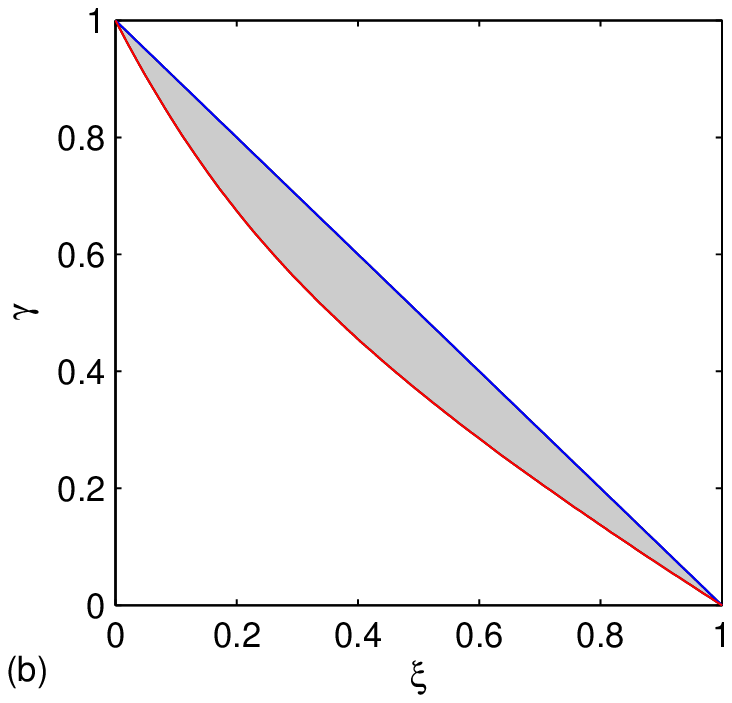}
\caption{\label{ab}(a) Comparison of nonlocalities before and after
distilling 'noisy correlated boxes' for the case
$d\rightarrow\infty$. In the figure, the gray surface represents the
set of initial boxes' CGLMP value and the colorful surface
represents the set of final boxes' CGLMP value. (b) The area our
protocol works for the case of $d\rightarrow\infty$. In the figure,
the blue line corresponds to the equation: $\xi+\gamma=1$ , and the
red line corresponds to the equation:
$4\xi^{2}+2\gamma^{2}+8\xi\gamma=4\xi+2\gamma$. In the shaded area
our protocol works well.}
\end{figure}

The CGLMP value of the initial box is $4\xi+2\gamma$. After
distillation, we get a box with CGLMP value
$(4+\frac{2}{d^2})\xi^2+(8-\frac{2}{d^2})\xi\gamma+\frac{2}{d^2}\xi+2\gamma^2$.
For a fixed $d$, it is easy to compare the strength of nonlocality
of the initial box and the final box. Consider the case
$d\rightarrow\infty$, the corresponding final box has the CGLMP
value of $4\xi^2+8\xi\gamma+2\gamma^2$. When the weight of the
non-local part $\xi$ and local part $\gamma$ range in the shaded
areas in Fig.\ref{ab}(b), the protocol works well for the
$d\rightarrow\infty$ case.

\section{Conclusions}\label{s5}
By adding the comparators into the wirings of the nonlocality
distillation, we realized the nonlocality distillation of
high-dimensional nonsignaling correlated nonlocal boxes. It shows
that high-dimensional nonlocality in general nonsignaling theory
measured by the CGLMP inequality also can be distilled by local
classical operation. An arbitrarily small violation of the CGLMP
inequality can be amplified to the asymptotic extremal violation
through a finite number of our distillation protocols. We used
different local compositions in the initial boxes, and found that
the initial boxes superposed by local deterministic boxes and
nonlocal extremal boxes can be distilled with a higher efficiency
than that of the case with the initial boxes being superposition of
local correlated boxes and nonlocal extremal boxes. The protocol
distilling different boxes with different efficiencies showed that
the nonlocality distillation is both wirings- and
components-selective.

Furthermore, in sec.\ref{s4}, we showed that our protocol also works
for the more general high-dimensional boxes. We also studied the
distillation of another general superposition: $P_{\xi,\gamma}^d=\xi
P_{\text{NL}}^d+\gamma P_{\text{Lc}}^d+(1-\xi-\gamma)\openone$, and
the nonlocality of the final one will be
$(4-\frac{2}{d^2})\xi^2+(6+\frac{2}{d}-\frac{2}{d^2})\xi\gamma+\frac{2}{d^2}\xi+2\gamma^2$
after the wirings in protocol A, which is similar to the protocol we
discussed in sec.\ref{s4}.

It is worth while to discuss some further problems. Are there
protocols distilling better for more high-dimensional boxes, e.g.
depth-3 protocols\cite{HR}? The closure of the high-dimensional
nonlocal polytope also can be discussed under the presented
distillation wirings. So far, all the nonlocality distillation
protocols are working for super-quantum resources and the wirings
are in the forms of local classical operations. Whether the local
quantum operations can amplify the nonlocalities of quantum states
still needs further studies. Finding the universal nonlocality
distillation protocol for arbitrary-partite arbitrary-dimensional
system needs more further research.

\begin{acknowledgments}

This work is supported by National Natural Science Foundation of China (NSFC) under Grants No. 10704001, No. 61073048, No.10905024 and 11005029, the Key Project of Chinese Ministry of Education.(No.210092), the Key Program of the Education Department of Anhui Province under Grants No. KJ2012A020, No. KJ2010A287, No. KJ2010A323, No. KJ2012B075 and No. 2010SQRL153ZD, the `211' Project of Anhui University, the Talent Foundation of Anhui University under Grant No.33190019, the personnel department of Anhui province and the research project of Lu'an city(2010LW027).
\end{acknowledgments}

\end{document}